\documentclass[preprint]{aastex}
\newcommand{\bm}[1]{{\mbox{\boldmath$#1$}}}
\shorttitle{Testing Propagation of Gravity}
\shortauthors{Kopeikin Sergei}
\begin{document}
\title{Testing Relativistic Effect of Propagation of Gravity by Very-Long Baseline Interferometry}
\author{Sergei M. Kopeikin}
\affil{Department of Physics and Astronomy, 322 Physics Building, University of
Missouri-Columbia, Columbia, MO 65211}
\email{kopeikins@missouri.edu}
\begin{abstract}\noindent
It is shown that the finite speed of gravity affects very-long
baseline interferometric observations of quasars during the time
of their line-of-sight close angular encounter with Jupiter. The next such event will
take place in 2002, September 8. The present {\it Letter} suggests a new
experimental test of general relativity in which the effect of propagation
of gravity can be directly measured by very-long baseline
interferometry as an excess time delay in addition to the
logarithmic Shapiro time delay (Shapiro, I. I., 1964,  Phys. Rev.
Lett., {\bf 13}, 789). 
\end{abstract}
\keywords{gravitation -- relativity -- techniques: interferometric
-- (galaxies:) quasars: individual (QSO J0842+1835)}
\newpage
\section{Introduction}
Experimental verifications of basic principles underlying Einstein's general relativity theory are important for fundamental physics. Numerous
tests of general relativity in the solar system \citep{1} and in
the binary pulsar PSR 1913+16 \citep{2} confirm its validity up to
the precision of 1\% or slightly better. 

It is worth
emphasizing that all of the Solar system tests of general relativity
have relied upon the Schwarzschild solution and could say nothing
about effects of retardation associated with the propagation of gravity.
Einstein's theory of general relativity predicts that if the second
time derivative of the quadrupole moment of a gravitating system
(e.g., a binary pulsar) is not zero,  the system emits gravitational
waves that travel outward at the speed of light.  Indirect
evidence for the existence of gravitational waves consists of
observations of the inspiraling orbits of binary pulsars \citep{2}.
Observations of binary pulsar inspiral for PSR 1913+16 agree at the level of $\sim 1\%$ with predictions based on the emission of energy by the binary in the form of (quadrupole)
gravitational radiation \citep{kop1985,gs1985,3}. This also provides
an indirect evidence that gravity waves must travel at the speed
of light \citep{3,1}. However, no
one has yet {\it directly} detected gravitational waves, let alone
measured their speed.

It is the purpose of this {\it Letter} to point out that observing propagation of light through the gravitational field
of the solar system can serve as a tool for measuring effects
associated with the finite speed of propagation of gravity. This is based on our previous papers \citep{kop1997,ksge,8}, where we have developed a
post-Minkowskian approach for solving the problem of propagation
of light through time-dependent
gravitational fields in the geometric optics approximation. This
approach is based on making use of the retarded {\it
Li\'enard-Wiechert-}type solutions of the linearized Einstein
equations and allows one to find a smooth analytic representation of
light-ray trajectory for arbitrary locations of the source of
light and observer without imposing any restrictions on the
motion of the light-ray deflecting bodies. We find that
electromagnetic signals interact with gravitating bodies only through the retarded gravitational field of the bodies. That is, if a
light-ray deflecting body moves with respect to a chosen
coordinate system the temporal variation in its gravity field must
take time to reach the electromagnetic signal in order to perturb
its trajectory. This observation constitutes the main idea of
the proposed VLBI test of the propagation of gravity
elaborated in the following sections of the present {\it Letter}. 

In 1964, I. I. \cite{9-a} suggested that the gravitational
deflection of light by the Sun --- one of the three classical effects
of general relativity  analyzed by Einstein --- could be measured
more accurately at radio wavelengths with interferometric
techniques than at visible wavelengths with the available optical
techniques. His idea led to stringent experimental limitations
on the parameter $\gamma$ of the parametrized post-Newtonian
formalism \citep{9}, thereby strongly restricting the number of viable
theories of gravity. Our
post-Minkowskian approach gives in the case of static
spherically-symmetric field the same result as predicted by
Shapiro. Furthermore, we are able to
calculate additional corrections to the Shapiro time delay related
to the non-stationarity of the gravitational field of the solar
system and associate them with the finite speed of propagation of
gravity. The largest contribution to the non-stationarity of
gravitational field of the solar system comes about via the orbital
motions of the most massive planets --- Jupiter and Saturn.
Therefore, it is reasonable to undertake an attempt to detect the
effect of propagation of gravity by observing very accurately
the deflection of light rays from a background source of light (quasar)
caused by the motion of Jupiter or Saturn. This is the essence of
our proposed VLBI test of general relativity discussed in
the present paper.

In section 2 we consider the basic formula for 
relativistic time delay in time-dependent gravitational fields.
Section 3 outlines the basic principles of measurement of the
effect of propagation of gravity in radio interferometric
experiments. Section 4 is dedicated to the description of the
proposed VLBI experiment and gives numerical estimates for the
Shapiro time delay in the field of Jupiter and for the effect of
propagation of gravity. Finally, in section 5 we discuss
the proposed VLBI experiment.

\section{Relativistic time delay in time-dependent gravitational fields}

Let us assume that the gravitational field is generated only by the
Solar system's bodies and there is a global four-dimensional
coordinate system $(t,{\it\bm x})$ with origin at the barycenter of
the Solar system. The total time of propagation of an
electromagnetic signal from the point $(t_0,{\it\bm x}_0)$
(quasar) to the point $(t,{\it\bm x})$ (observer) is given by the
expression
\begin{equation}
\label{1}
t-t_0=\frac{1}{c}|{\it\bm x}-{\it\bm x}_0|+\Delta(t,t_0)\;.
\end{equation}
Here, $|{\it\bm x}-{\it\bm x}_0|$ is the usual Euclidean distance
between the points of emission, ${\it\bm x}_0$, and observation,
${\it\bm x}$, of the photon, and $\Delta(t,t_0)$ is the relativistic
time delay produced by the gravitational field of the moving
gravitating bodies of the Solar system. The function $\Delta(t,t_0)$
is given by \citep{8}
\begin{eqnarray}
\label{3} \Delta(t,t_0)&=&{2G\over c^2}\sum_{a=1}^N
m_a\;\displaystyle{\int^{s}_{s_{0}}}\frac{\left(1-c^{-1}{\it\bm
k}{\bm\cdot}{\it\bm v}_a(\zeta)\right)^2}{\sqrt{1-c^{-2}v^2_a(\zeta)}}
\frac{d\zeta}{c t^{\ast}+{\it\bm
k}{\bm\cdot}{\it\bm x}_a(\zeta)-c \zeta}\;,
\end{eqnarray}
where $m_a$ is the mass of the $a$th body, $t^*$ is the time of the closest approach of electromagnetic
signal to the barycenter of the Solar system \footnote{The time
$t^*$ is used in calculations as a mathematical tool only.}, ${\it\bm x}_a(t)$ are coordinates of the
$a$th body, ${\it\bm v}_a(t)=d{\it\bm x}_a(t)/dt$ is the
(non-constant) velocity of the $a$th light-ray deflecting body,
${\it\bm k}$ is the unit vector from the point of emission to the
point of observation, $s$ is a retarded time obtained by solving
the gravitational null cone equation for the time of observation
of photon $t$, and $s_0$ is found by solving the same equation
written down for the time of emission of the photon $t_0$
\begin{equation}
\label{4}
s+{1\over c}|{\it\bm x}-{\it\bm x}_a(s)|=t\;,\qquad\qquad
s_0+{1\over c}|{\it\bm x}_0-{\it\bm x}_a(s_0)|=t_0\;.
\end{equation}

It is remarkable that formula (\ref{3}) does not impose any
restriction on the motion of bodies since we did not use any
explicit solutions of their equations of motion while calculating the light ray propagation. The formula (\ref{3}) can be further simplified by performing integration by parts which yields \footnote{After integration we omitted for simplicity terms of order $v_a^2/c^2$.}
\begin{eqnarray}
\label{5} \Delta(t,t_0)&=&-\sum_{a=1}^N {2Gm_a\over
c^3} \biggl\{\ln\left[\frac{r_a(s)-{\it\bm k}{\bm\cdot}{\it\bm
r}_a(s)} {r_{0a}(s_{0})-{\it\bm k}{\bm\cdot}{\it\bm
r}_{0a}(s_{0})}\right]-{{\it\bm
k}{\bm\cdot}{\it\bm v}_a(s)\over c}\ln\biggl[r_a(s)-{\it\bm
k}{\bm\cdot}{\it\bm r}_a(s)\biggr]+
\\\nonumber\\&&\nonumber\hspace{-1cm}
{{\it\bm k}{\bm\cdot}{\it\bm v}_{a0}(s_0)\over c}\ln\biggl[r_{0a}(s_0)-{\it\bm
k}{\bm\cdot}{\it\bm r}_{0a}(s_0)\biggr]- \int^s_{s_{0}}
\ln\biggl[c t^{\ast}+{\it\bm k}{\bm\cdot} {\it\bm
x}_a(\zeta)-c \zeta\biggr]{{\it\bm k}{\bm\cdot}{\it\bm a}_a(\zeta)\over
c}\;d\zeta\biggr\}\;,
\end{eqnarray}
where ${\it\bm r}_a={\it\bm x}-{\it\bm x}_a(s)$, ${\it\bm
r}_{0a}={\it\bm x}_0-{\it\bm x}_{a}(s_0)$, $r_a=|{\it\bm r}_a|$,
$r_{0a}=|{\it\bm r}_{0a}|$, ${\it\bm v}_a=\dot{\it\bm x}_a(s)$,
${\it\bm v}_{a0}=\dot{\it\bm x}_{a}(s_0)$, ${\it\bm
a}_a=\dot{\it\bm v}_a(s)$, and the retarded times $s$ and $s_0$
are calculated from the gravitational null-cone equations
(\ref{4}).

Our expression (\ref{5}) vastly extends the region of
applicability of Shapiro's work \citep{9} for it is valid
for the case of arbitrary-moving bodies whereas the calculations
by Shapiro had been done only for the case of static
Schwarzschild field.  Earlier work on
improving the Schwarzschild approximation in the case of light passing
through the gravitational field of moving bodies should be noted \citep{13,12,
kk92,14}; however, our expression (\ref{5}) is much more general and
completely solves the problem. We emphasize that solution (\ref{5}) is valid
everywhere both inside and outside of the Solar system at
arbitrary distances including its near, intermediate, and wave zones, and, for this reason, represents a smooth
analytic solution of the equations of light-ray geodesics for
arbitrarily located source of light, observer, and the system of the
light-ray deflecting bodies. However, the most important observation
is that our formula (\ref{5}) allows one to single out the retardation
effects uniquely associated with the finite speed of propagation
of gravity and indicates that for the correct calculation of the relativistic time delay positions of the gravitating bodies
must be taken at the retarded times $s_0$ and $s$ corresponding to
the instants of emission $t_0$ and observation $t$ of the
electromagnetic signal 

\section{Principles of Measurement of the Propagation of Gravity by VLBI}

Very Long Baseline Interferometry (VLBI) has recently reached a
precision in measurements of differential phase delay of order $10^{-12}$ second (ps), which makes it presently the most
accurate technique for measuring relativistic time delay \citep{bc,hks}. In order to discuss how to measure the relativistic effect of propagation
gravity by VLBI one transforms the formula (\ref{5}) to the form used in VLBI analysis and
called the differential VLBI time delay \citep{16}. In
doing this conversion, we note that formula (\ref{5}) is also valid in scalar-tensor
theories after replacement of the universal gravitational
constant $G\rightarrow G(1+\gamma)/2$, where $\gamma$ is one of
the parameters of the PPN formalism \citep{1}.

Let us assume now that there are two earth-based VLBI stations and
the plane front of electromagnetic waves from a quasar propagates towards the Earth. Taking two rays from the wave front
and subtracting equation (\ref{1}) for the
first light ray from that for the second ray yields
\begin{equation}
\label{6}
t_2-t_1={1\over c}|{\it\bm x}_2-{\it\bm x}_0|-{1\over c}|
{\it\bm x}_1-{\it\bm x}_0|+\Delta(t_1,t_2)\;,
\end{equation}
where $\Delta(t_1,t_2)\equiv\Delta(t_2,t_0)-\Delta(t_1,t_0)$. Here
$t_1$ and $t_2$ are coordinate times of arrival of the
electromagnetic signal from the source of electromagnetic waves
(quasar) to the first and second stations, respectively, and ${\it\bm
x}_1$ and ${\it\bm x}_2$ are spatial coordinates of the first and second
VLBI stations with respect to the barycentric frame of the solar
system that is chosen as the primary non-rotating reference frame
\citep{16}.

The difference $\Delta(t_1,t_2)$ 
is obtained from the main formula (\ref{5}) after long and tedious
calculations and analysis of residual terms. It can be
proved \citep{8} that with accuracy better than 1 ps,
\begin{eqnarray}
\label{7} \Delta(t_1,t_2)&=&(1+\gamma)\sum_{a=1}^N {Gm_a\over
c^3}\biggl(1+{{\it\bm K}{\bm\cdot}{\it\bm v}_a(s_1)\over c}\biggr)\;
\ln\frac{r_{1a}(s_1)+{\it\bm K}{\bm\cdot}{\it\bm r}_{1a}(s_1)}
{r_{2a}(s_2)+{\it\bm K}{\bm\cdot}{\it\bm r}_{2a}(s_2)}\;,
\end{eqnarray}
where ${\it\bm K}$ is the unit vector from the
barycenter of the solar system to the quasar; ${\it\bm v}_a(s_1)$ is
the velocity of the $a$th gravitating body at time $s_1$, $r_{1a}=|{\it\bm r}_{1a}|$,
$r_{2a}=|{\it\bm r}_{2a}|$, ${\it\bm r}_{1a}(s_1)={\it\bm
x}_{1}(t_1)-{\it\bm x}_{a}(s_1)$ and ${\it\bm r}_{2a}(s_2)={\it\bm
x}_{2}(t_2)-{\it\bm x}_{a}(s_2)\;$; moreover, the retarded times $s_1$ and
$s_2$ are calculated according to 
\begin{equation}
\label{8}
s_1=t_1-{1\over c}|{\it\bm x}_1(t_1)-{\it\bm x}_a(s_1)|\;,\qquad\qquad\qquad
s_2=t_2-{1\over c}|{\it\bm x}_2(t_2)-{\it\bm x}_a(s_2)|\;.
\end{equation}

The effect of propagation of gravity appears in equation (\ref{7})
as a displacement of the light-ray deflecting bodies from their
present to retarded positions. We note that velocities of the
gravitating bodies ${\it\bm v}_a$ are small with respect to the
speed of gravity and the time taken by the light ray to reach the observer
after passing by the gravitating body is much smaller than its
orbital period. Thus, one can expand the retarded positions
${\it\bm x}_a(s_i)\;(i=1,2)$ of the bodies in Taylor series
around their present positions ${\it\bm x}_a(t_i)$ taken at the arrival
times $t_i\;(i=1,2)$. Neglecting all terms quadratic with respect
to velocities of the gravitating bodies and/or proportional to
their accelerations one gets
\begin{equation}
\label{9} \Delta(t_1,t_2)=(1+\gamma)\sum_{a=1}^N {Gm_a\over
c^3}\biggl[\left(1+{{\it\bm K}{\bm\cdot}{\it\bm v}_a\over
c}\right) \ln\frac{r_{1a}+{\it\bm K}{\bm\cdot}{\it\bm r}_{1a}}
{r_{2a}+{\it\bm K}{\bm\cdot}{\it\bm r}_{2a}}- {{\it\bm
B}{\bm\cdot}{\it\bm v}_a+({\it\bm K}{\bm\cdot}{\it\bm
v}_{a})({\it\bm N}_{1a}{\bm\cdot}{\it\bm B})\over c(r_{1a}+{\it\bm
K}{\bm\cdot}{\it\bm r}_{1a})}\biggr]\;,
\end{equation}
where ${\it\bm B}={\it\bm x}_2(t_1)-{\it\bm x}_1(t_1)$ is the
baseline between the two VLBI stations, ${\it\bm r}_{ia}={\it\bm
x}_{i}(t_i)-{\it\bm x}_{a}(t_i)$ is the difference of coordinates
of the $i$th VLBI station and $a$th gravitating body taken at
the time of arrival of the radio signal to the $i$th station,
${\it\bm N}_{1a}={\it\bm r}_{1a}/r_{1a}$, ${\it\bm
v}_a\equiv{\it\bm v}_a(t_1)$ and all quantities in the second term
on the right side of equation (\ref{9}) are also evaluated at
time $t_1$.

The logarithmic term on the right side of equation (\ref{9})
was first discovered by \cite{9}. However, the finite speed of
propagation of gravity leads to the appearance of an additional (second)
term on the right side of equation (\ref{9}) as well. Had the speed of gravity
been set equal to infinity as it is done in Newtonian
gravitation, the second term on the right side of
equation (\ref{9}) would be identically equal to zero.
Experimental confirmation of the existence of such a term would directly
prove that gravity propagates with a finite speed.

\section{The Proposed Radio Interferometric Experiment}

In 2002, on September 8 Jupiter will pass at the angular distance of about
3.7 arcminutes from the quasar QSO J0842+1835, making an ideal
celestial configuration for measuring the speed of propagation of
gravity by using the differential VLBI technique. During the passage
of Jupiter near the quasar line of sight, the time-dependent impact parameter
$|{\bm{\xi}}|$ of the light ray from the quasar with respect to
Jupiter will be always small as compared with the 
Earth-Jupiter distance, $r_{\oplus J}$, which is approximately equal to 6
astronomical units. Hence, it is convenient to introduce the unit
vector ${\it\bm n}$ along the direction of the impact parameter
according to the definition ${\bm{\xi}}=|{\bm{\xi}}|\;{\it\bm n}$
such that
\begin{equation}
\label{imp} {\it\bm N}_{1J}=-\cos\theta\;{\it\bm
K}+\sin\theta\;{\it\bm{n}}\;,
\end{equation}
where the subscript $J$ refers to Jupiter and $\theta\simeq
|{\bm{\xi}}|/r_{\oplus J}$ is the (small) angle between the
undisturbed geometric positions of the quasar and Jupiter. The baseline ${\it\bm B}$ during the time of the
experiment will be much smaller than $|{\bm{\xi}}|$; therefore, the following
expansions are valid
\begin{eqnarray}
\label{10} {\it\bm N}_{1J}&=&-\left(1-{\theta^2\over
2}\right){\it\bm K}+\theta\;{\it\bm{n}}+O(\theta^3)\;,
\\\label{11}
r_{2a}+{\it\bm K}{\bm\cdot}{\it\bm r}_{2a}&=&r_{1a}+{\it\bm
K}{\bm\cdot}{\it\bm r}_{1a}+{\it\bm N}_{1a}{\bm\cdot}{\it\bm B}+
{\it\bm K}{\bm\cdot}{\it\bm B}+O({\it\bm B}^2)\;.
\end{eqnarray}

Other than Jupiter, the Earth and the Sun also contribute significantly to the
gravitational time delay and must be included in the data
processing algorithm in order to extract the effect of propagation
of gravity unambiguously and to measure its speed. Accounting for
this fact and making use of the expansions (\ref{10})--(\ref{11}),
we can recast formula (\ref{9}) into the following convenient
form
\begin{equation}
\label{12}\Delta(t_1,t_2)=\Delta_\oplus+\Delta_\odot-(1+\gamma)\biggl[{2GM_J\over
c^3r_{1J}}{{\it\bm n}{\bm\cdot}{\it\bm B}\over
\theta}+(1+\delta){2GM_J\over c^4r_{1J}}{{\it\bm
B}{\bm\cdot}{\it\bm v}_J-({\it\bm K}{\bm\cdot}{\it\bm
v}_J)({\it\bm K}{\bm\cdot}{\it\bm B})\over \theta^2}\;\biggr]\;,
\end{equation}
where we have introduced a new phenomenological parameter
$\delta$ parameterizing the effect of propagation of gravity in data processing algorithms. Thus, we emphasize that there are two
relativistic parameters to be measured in the VLBI experiment in order to test the validity
of general relativity theory --- the PPN parameter $\gamma$ and the
gravity propagation parameter $\delta$. The best experimental
measurement of the parameter $\gamma$ is due to Lebach et al. (1995), who obtained $\gamma=0.9996\pm 0.0017$ in 
excellent agreement with general
relativity. The primary goal of the new experimental test of
general relativity proposed in the present {\it Letter} is to set {\it direct observational} limits on the
parameter $\delta$ which will measure the effect of retardation in propagation of gravity by the moving Jupiter. According to the Einstein
theory of relativity one must expect that the
numerical value of the parameter $\delta$ must be equal to zero.

We emphasize that our consideration is fully based on the Einstein theory of general relativity in which the speed of propagation of gravity is equal to the speed of light in vacuum. \cite{w} analyzed the case when the two speeds are not equal and showed that the resulting theory would have a non-zero value of the PPN parameter $\alpha_2=(c/c_g)^2-1$, where $c_g$ is the value of the propagation speed of gravity in the rest frame of
the universe.\footnote{We would like to emphasize that our parameter $\delta$ is just a fiting parameter and has no any relation to the PPN parameter $\alpha_2$ as we work in the framework of general relativity where $c_g=c$.} This parameter has been strongly bounded by consideration of various
``preferred frame effects''. The present best limit \citep{n87} is $|\alpha_2| < 4 \times 10^{-7}$. The experiment proposed in the present {\it Letter} is not designed to compete with this tight bound, rather its main purpose is to measure the retardation effect (parameter $\delta$) associated with finite speed of propagation of gravity which was never observed before.  

The first term on the right side of equation (\ref{12})
describes the Shapiro time delay due to the
gravitational field of the Earth. It can reach 21 ps for a
baseline of $B\simeq 6000$ km. The second term on the right hand side of
equation (\ref{12}) describes the Shapiro time
delay due to the Sun. It can vary (for $B\simeq 6000$ km) from $17\times
10^4$ ps for the light ray grazing the Sun's limb to only 17 ps
when the direction to the source of light is opposite to the Sun
\citep{12}. The third term on the right side of equation
(\ref{12}) is the standard Shapiro time delay due to the static
gravitational field of Jupiter. Finally, the fourth term on the
right side of equation (\ref{12}) is the time delay caused by
the effect of propagation of gravity. Numerical estimates of the
Shapiro time delay $\Delta_J$ and the gravity propagation
time delay $\Delta_{JP}$ caused by Jupiter can be obtained
directly from equation (\ref{12}). One finds for $\gamma=1$ and $\delta=0$, 
\begin{equation}
\label{14} \Delta_J\simeq\alpha_\odot\left({R_\odot\over
R_J}\right)\left({M_J\over
M_\odot}\right)\left({\theta_J\over\theta}\right)\left({R_\oplus\over
c}\right)\;,\qquad\qquad
\Delta_{JP}\simeq\left({\Delta_J\over\theta}\right)\left({v_J\over
c }\right)\;,
\end{equation}
where $\alpha_\odot=1.75''$ is the relativistic deflection of
light by the Sun for the case of light grazing the Sun's limb,
$R_\odot$ is the radius of the Sun, $R_J$ is the radius of Jupiter, $R_\oplus$ is the radius of the Earth, 
$\theta_J=0.27'$ is the visible angular radius of Jupiter at the
distance of $r_{\oplus J}=6$ AU, $\theta=3.7'$ is the minimum angular distance from
Jupiter to the quasar in 2002, on September 8, and 
$v_J$ is the orbital speed of
Jupiter. 

It is worth emphasizing that $\Delta_J\sim\theta^{-1}$, while the
gravity propagation time delay $\Delta_{JP}\sim\theta^{-2}$. For
this reason, the effect of propagation of gravity is not strongly
suppressed by the presence in $\Delta_{JP}$ of the small factor
$v_J/c=4.5\times 10^{-5}$, and, it turns out to be 
measurable using VLBI techniques. Indeed, using the
numerical values for the parameters in equation
(\ref{14}) for the case of the Jupiter-quasar "encounter" in 2002,
September 8, we obtain ($B\simeq R_\oplus$)
\begin{equation}
\alpha_J\simeq 1.26\;\; \mbox{mas}\;,\qquad\quad\Delta_J\simeq 137\;
\mbox{ps}\;,\qquad\quad\Delta_{JP}\simeq 6 \;\mbox{ps}\;,
\end{equation}
so that $\Delta_{JP}/\Delta_{J}\simeq 0.04$. Hence, the correction due to the propagation of gravity would be $\sim 4\%$ of the Shapiro time delay. With the present differential VLBI accuracy it appears that the new effect ($\Delta_{JP}\simeq 6$ ps) is measurable with careful
observations. Indeed, one can achieve a phase-referencing VLBI accuracy of a
few degrees of phase (averaged over a 10-hour observation) and can
detect relative astrometric position changes of less than 0.010 mas ($\sim 1$ ps) with
the existing VLBA and Effelsberg radio antennas (E. Fomalont, private communication).

Formula (\ref{14}) also explains why it is not so effective to observe the effect of the propagation of gravity in the field of the Sun. It turns out that for the case of the Sun the angle $\theta$ can not be made as small as in the case of Jupiter, because the Sun is closer to the Earth and, in addition, its radius is ten times larger than that of Jupiter. This makes the effect of the propagation of gravity in the field of the Sun $\leq 10$ ps for the light ray grazing the Sun's limb.

\section{Discussion of the Proposed VLBI Experiment}

The effect of propagation of gravity appears in addition to the
logarithmic Shapiro time delay as an excess delay of $\sim 4\%$. The Shapiro time delay $\Delta_J$ caused
by the static part of the gravitational field of Jupiter was first
measured in 1991 \citep{28}. Klioner (1991) used the post-Newtonian expression for the metric tensor under the assumption that the planets of the solar system move uniformly along straight lines. He obtained the same result for $\Delta_{JP}$ formally; however, due to the specific assumptions used for the derivation of the near field post-Newtonian metric, he was not able to interpret it properly as the effect caused by the finite speed of propagation of gravity. This is because there was no proof that the solution of the light geodesic equations obtained on the basis of the near field post-Newtonian metric can be smoothly matched with the far field metric describing emission gravitational waves by the orbital motion of planets in the Solar system. The present {\it Letter} provides evidence that this near field post-Newtonian metric directly depends on the speed of propagation of gravity in general relativity. 

Other authors \citep{13,12,14} have also attempted to derive more exact expressions for the Shapiro effect in the
case of time-dependent gravitational field of the Solar system.
These authors used the post-Newtonian metric for the calculation of the Shapiro effect with positions of bodies taken at the time of the closest approach $t^*_a$ of a radio signal to the body
\begin{equation}
\label{tca}
t^*_a=t-{1\over c}{\it\bm k}{\bm\cdot}\left({\it\bm x}-{\it\bm
x}_a\right)\;.
\end{equation}
The instant of time $t^*_a$ is numerically close to the retarded
time $s$ calculated as the solution of the gravitational null cone
equation (\ref{4}) and, for this reason, makes a good
approximation for the calculation of the Shapiro time delay. However,
in the general case when the impact parameter is large, $t^*_a$ and $s$ can have very different
numerical values. Indeed, the physical meanings
of the time of the closest approach $t^*_a$ and the retarded time
$s$ are crucially different. As it is seen from its definition
(\ref{tca}), the time $t^*_a$ is calculated only approximately
using the finite speed of propagation of {\it light}, while the
retarded time $s$ emerges in our more exact post-Minkowskian analytic calculations (\ref{5}) due to
the finite speed of propagation of {\it gravity}. It is
because of this significant difference that the VLBI experiment in
2002, September 8 will probe the non-trivial effect of the
propagation of gravity.

It is also useful to see how the effect of the propagation of gravity can be explained qualitatively without doing lengthy calculations. For this purpose, we note that equation (\ref{7}) indicates that the Shapiro time delay is actually a function of the retarded time $\Delta_J(s)$. Expanding this function in Taylor series around the time of observation $t$, one finds
\begin{eqnarray}
\label{exp}\nonumber
\Delta_J(s)&=&\Delta_J(t)+\dot{\Delta}_J(t)(s-t)+...\\&\simeq &\Delta_J(t)+\Delta_J(t)\left({\dot{\theta}\over\theta}\right)\left({r_{\oplus J}\over c}\right)+...=\Delta_J(t)+\Delta_{JP}(t)+...\;,
\end{eqnarray}    
where $\dot{\Delta}_J(t)\simeq -(\dot{\theta}/\theta)\Delta_J$ using equation (\ref{14}), $s-t\simeq -r_{\oplus J}/c$ using equation (\ref{4}),  and $\Delta_{JP}$ is given by the approximate expression (\ref{14}).

One can see that the delay $\Delta_{JP}$ due to the finite speed of propagation of gravity can become comparable with the standard (instantaneous) prediction for the Shapiro time delay $\Delta_{J}(t)$ in case $\theta\simeq v_J/c$. This can never happen for Jupiter and the other solar system bodies, but can be interesting for VLBI experiments if one were able to find a binary star close to the line of sight of a quasar. Orbital motion of the binary star would cause periodic modulation of the Shapiro delay in the time of propagation of light from the quasar to the observer, which can be used for the determination of the effect of propagation of gravity outside the solar system \citep{ksge,kg}. 
It is worth noting that the effect of propagation of gravity can also be  studied in the timing of binary pulsars with nearly edgewise orbits and by the doppler tracking of spacecraft in deep space. 

\acknowledgments

We are indebted to Prof. B. Mashhoon who participated in numerous
scientific discussions, carefully read the manuscript and made
many essential suggestions for its improvement. We are grateful to Dr. S.M. Kudryavtsev for 
calculating the date of the Jupiter's encounter with QSO J0842+1835.
Thanks are also due to Dr. E. Fomalont, Dr. L. Petrov, Prof. K. Nordtvedt, Prof. J. Burns, Prof. C. R. Gwinn, and Prof. O. Sovers for interesting and stimulating discussions and suggestions.

\end{document}